\documentclass[aps,floatfix,twocolumn,pra,nofootinbib,superscriptaddress,longbibliography]{revtex4-2}
\usepackage{array}
\usepackage{amsmath,amsfonts,amssymb,stackrel}
\usepackage{xurl}
\PassOptionsToPackage{hyphens}{url}
\usepackage[breaklinks=true]{hyperref}
\hypersetup{colorlinks=true}
\hypersetup{urlcolor=blue}
\hypersetup{citecolor=black}
\hypersetup{menucolor=black}
\hypersetup{linkcolor=black}
\usepackage{cleveref}
\usepackage{float,cancel,makecell}
\usepackage[T1]{fontenc}
\usepackage[utf8]{inputenc}
\usepackage{tikz}

\newcommand{\overlinef}[1]{\overline{\vphantom{\pp'}#1}}

\newcommand{\be}{\begin{equation}}
\newcommand{\ee}{\end{equation}}
\newcommand{\bea}{\begin{eqnarray}}
\newcommand{\eea}{\end{eqnarray}}

\newcommand{\meanv}[1]{\langle #1 \rangle}

\newcommand{\meanvlr}[1]{\left\langle #1 \right\rangle}

\newcommand{\bb}[1]{\left( #1 \right)}

\newcommand{\bbcro}[1]{\left[ #1 \right]}
\newcommand{\bbcror}[1]{\left. #1 \right]}
\newcommand{\bbcrol}[1]{\left[ #1 \right.}

\newcommand{\ii}{\textrm{i}}
\newcommand{\dd}{\textrm{d}}
\newcommand{\eee}{\textrm{e}}
\newcommand{\rr}{\textbf{r}}

\newcommand{\qq}{\textbf{q}}

\newcommand{\pp}{\textbf{p}}

\def\bleu{\color{blue}}

\newcommand{\ket}[1]{|#1\rangle}
\newcommand{\bra}[1]{\langle #1|}

\newcommand{\kF}{k_{\rm F}}

\newcommand{\TF}{T_{\rm F}}

\newcommand{\pF}{p_{\rm F}}
\newcommand{\vF}{v_{\rm F}}

\newcommand{\upa}{\uparrow}
\newcommand{\dwa}{\downarrow}

\newcommand{\eqqref}[1]{Eq.~\eqref{#1}}
\newcommand{\eqqrefs}[2]{Eqs.~\eqref{#1}--\eqref{#2}}

\newcommand{\braket}[2]{\langle #1|#2\rangle}

\usepackage{tikz}
\usetikzlibrary{shapes.geometric, positioning, shadings, arrows.meta}
\usetikzlibrary{calc,patterns,decorations.pathmorphing,decorations.markings}

\tikzset{->-/.style={decoration={
  markings,
  mark=at position .5 with {\arrow{>}}},postaction={decorate}}}
\tikzset{-<-/.style={decoration={
  markings,
  mark=at position .5 with {\arrow{<}}},postaction={decorate}}}
\tikzset{phantom->-/.style={decoration={
  markings,
  mark=at position .5 with {\arrow[scale=2]{>}}},postaction={decorate}}}

\tikzset
  {
    ,my bubble/.style = 
      {
        ,draw=#1!70
        ,fill=#1!10
        ,ellipse
        ,inner sep=1pt
        ,minimum width=2em
        ,minimum height=2em
        ,align=center
      }
    ,my end/.style =
      {
        ,draw=#1!70
        ,top color=#1!10
        ,bottom color=#1!50
        ,minimum height=6em
        ,text width=6em
        ,inner sep=0pt
        ,align=center
      }
    ,my arrow/.style =
      {
        ,>=Stealth
        ,->
        ,draw=black
      }
  }
  
\tikzset{serpent/.style={decoration={snake},postaction={decorate}}}

\begin{document}
\title{Exact perturbative expansion of the transport coefficients of a normal low-temperature Fermi gas with contact interactions}

\newcommand{\hkre}[1]{{\bleu #1}}

\author{Pierre-Louis Taillat}
\affiliation{Laboratoire de Physique Théorique de la Matière Condensée,
Sorbonne Université, CNRS, 75005, Paris, France}
\author{Hadrien Kurkjian}
\email{hadrien.kurkjian@cnrs.fr}
\affiliation{Laboratoire de Physique Théorique de la Matière Condensée,
Sorbonne Université, CNRS, 75005, Paris, France}

\begin{abstract}
We compute the shear viscosity, thermal conductivity and spin diffusivity of a
Fermi gas with short-range interactions in the Fermi liquid regime of the normal phase,
that is at temperatures $T$ much lower than the Fermi temperature $T_{\rm F}$ and larger
than the superfluid critical temperature $T_c$. 
In line with recent advances in the precision of cold atom experiments, we provide
exact results up to second-order in the interaction strength. 
We extend the Landau-Salpeter equation to compute the 
collision amplitude beyond the forward-scattering limit, covering all collisions on the Fermi surface.
We treat the collision kernel exactly, leading to significant 
corrections beyond relaxation-time or variational approximations.
The transport coefficients, as functions of the $s$-wave scattering length $a$ and Fermi
wavenumber $k_{\rm F}$,
follow $(1+\gamma k_{\rm F}a)/a^2$ up to corrections
of order $O(a^0)$,
with a positive coefficient
$\gamma$ for the viscosity and negative one for the thermal
conductivity and spin diffusivity.
\end{abstract}
\maketitle

The dynamics of fermionic matter is typically
a strongly-correlated many-body phenomenon, very challenging
to harness using controlled analytical approaches. In the general case, 
the absence of separation between few-body and many-body timescales 
invalidates standard kinetic assumptions like the “molecular chaos” \cite{SmithJensen}
and few-body correlations cannot be described  as collisions, in the sense of isolated Markovian events.
A more tractable scenario arises when the system obeys a kinetic equation \cite{Bonitz1998};
its many-body dynamics can be captured not only by a one-body phase-space 
density but also by an instantaneous differential 
equation. The hydrodynamic regime applies to 
the largest time and spatial scales when
the whole transport dynamics folds onto a system of Navier-Stokes equations 
on the conserved quantities, just like in classical fluids.
Dissipative properties are then entirely determined by transport 
coefficients, which quantify the impact of collisions on 
the conserved quantities. However, computing these coefficients
from first principles remains a longstanding challenge.

The kinetic description applies to Fermi systems only in specific limits: the 
weakly-interacting regime \cite{Vichi2000,Nikuni2009},
the high-temperature virial regime  \cite{Schafer2010,Nishida2019,Enss2023b}, 
and finally at low temperature within the Fermi liquid framework.
In the latter case, a low-energy effective description  of the 
system as a dilute gas of quasiparticles \cite{Landau1956} is necessary to truncate the hierarchy
of correlations. Even when quasiparticle interactions are large, 
the smallness of the temperature $T$ ensures that the
excited quasiparticles remain dilute with respect to the scattering crosssection, 
maintaining the separation of timescales.
The dynamics of a Fermi liquid is then determined by two
key functions describing the low-energy physics near the Fermi surface:  
the quasiparticle interaction function $f$ and the collision probability $W$. 
While Landau's Bethe-Salpeter equation \cite{Landau1959}
relates the forward-scattering limit of $W$ to $f$,
this is not sufficient for
a quantitative determination of the transport properties, which requires instead
the general expression of $W$ for four non-collinear momenta near the Fermi surface
\cite{BaymPethick}.

Fermi liquids are found in a variety of contexts,
from nuclear physics \cite{Ripka1985,Polls2016} to condensed 
matter \cite{FetterWalecka} or liquid Helium 3 \cite{BaymPethick,Wheatley1966}.
In most cases, theoretical predictions, 
restricted by the lack of knowledge on $W$,
remain qualitative.
Ultracold fermionic gases are a remarkable exception \cite{Zwerger2012}.
Atomic interactions are amenable to a unique
parameter, the $s$-wave scattering length $a$, enabling
microscopic calculations of $f$ \cite{LandauLipschitzVol9} and $W$.
Ultracold gases offer very clean and controllable platforms,
where one controls for instance the trapping geometry \cite{Hadzibabic2013,Hadzibabic2021} or 
the interaction strength \cite{Tiesinga2010}. Quantitative measurements of transport properties,
in particular of transport coefficients, are already accessible both 
in harmonic traps \cite{Thomas2008,Grimm2008,Zwierlein2011,Zwierlein2011Universal,Kohl2012,Thomas2015,Thomas2024}, 
homogeneous systems \cite{Zwierlein2019,yaleexp} and optical lattices \cite{Esslinger2007}.
In the regime of weak and negative scattering length, the smallness
of the critical temperature $T_c$ opens a regime where the gas is
both normal and quantum degenerate; it should thus behave as an excellent Fermi liquid.

Although this regime could provide a precious
quantitative test of Fermi liquid theory, and appears 
as a prerequisite to understand more difficult situations, such as
the dynamics near the phase transition \cite{Emery1976,Reppy1978,sergheitc,Thomas2015,Sauls2022} 
or at unitarity \cite{Schafer2010,Zwerger2011,Nishida2019,Enss2023b,Thywissen2018},
existing theoretical predictions of its properties are still
qualitative \cite{Smith2005,Nikuni2009,Bruun2011,Hofmann2025}, relying on uncontrolled
approximations for either the collision probability, or the solution
of the transport equation. A notable exception is the set of studies 
by Rainwater and M\"ohling \cite{Rainwater1975,Mohling1976}
which addresses the so-called ``dilute hard-sphere Fermi gas''
as an academic example in the context of nuclear physics \cite{thesemecca,Polls2016}.

Here, we present a controlled calculation of the three transport coefficients 
-- shear viscosity $\eta$, thermal conductivity $\kappa$
and spin diffusivity $D$ -- of the low-temperature unpolarized Fermi gas, 
including corrections linear in $k_{\rm F}a$.
Our perturbative expansion is exact,
avoiding the forward-scattering approximation 
on the collision amplitude, and treating
the collision kernel exactly \cite{Hofmann2023,Hofmann2025},
 without a relaxation-time \cite{Khalatnikov1957kinetic}, or a variational approximation
\cite{Smith2005}.
Our solution of the transport equation in the hydrodynamic regime
based on the expansion of the quasiparticle distribution over orthogonal polynomials \cite{Hofmann2023,disphydro},
also retains the contribution of the quantum force
to the viscosity and spin diffusivity, unlike the standard 
result from Refs.~\cite{Sykes1968,Wilkins1968}.

Notably, our result on the shear viscosity contradicts
the result by Rainwater and M\"ohling~\cite{Mohling1976} and matches well
the value measured in Yale at $k_{\rm F}a=-0.67$ \cite{yaleexp}
(where $k_{\rm F}$ is the Fermi wavenumber).
The coefficient of the correction linear in $a$ to the viscosity $\eta$ has a positive sign, which
(for $a<0$) accelerates the decreasing behavior in $1/a^2$ to leading order.
Conversely, for $\kappa$ and $D$  the correction has a negative
sign, which slows the decrease.

\textit{Quasiparticles in a weakly-interacting Fermi gas}---
We consider a gas of spin-$1/2$ fermions of mass $m$ and chemical 
potential $\mu$ trapped in a cubic volume $L^3$ and interacting via 
contact interactions, characterized by the bare coupling constant $g_0$. 
The Fermi momentum $\pF$ and equilibrium density $\rho_0=\pF^3/3\pi^2$ are related
to $\mu$ by the equation of state.
The Hamiltonian of the system reads:
\begin{eqnarray}
\hat{H}&=&\hat{H}_0+\hat{V}, \qquad \hat{H}_0=\sum_{\pp\in\mathcal{D},\sigma} 
{\frac{p^2}{2m}} \hat a_{\pp\sigma}^\dagger \hat a_{\pp\sigma} \\
\hat{V}&=&\frac{g_0}{L^3}\sum_{\pp_1,\pp_2,\pp_3,\pp_4\in
\mathcal{D}} \delta_{\pp_1+\pp_2}^{\pp_3+\pp_4} 
\hat a_{\pp_1\upa}^\dagger \hat a_{\pp_2\dwa}^\dagger \hat a_{\pp_3\dwa} \hat a_{\pp_4\upa}
\end{eqnarray}
where $\hat a_{\pp\sigma}$ annihilates a fermion of momentum $\pp$ and spin 
$\sigma$. To regularize the UV divergences, we have discretized real space \cite{Varenna} into a cubic lattice of step $l$, thereby restricting the set of momenta $\pp$ to 
$\mathcal{D}=(2\pi \mathbb{Z}/L)^3\ \cap\ [-\pi/l,\pi/l)^3$. Solving the two-body problem, we express $g_0$ in terms of the scattering length $a$
\begin{equation}
\frac{1}{g_0}=\frac{1}{g}-\int_{[-\pi/l,\pi/l)^3}\frac{\dd^3 p}{(2\pi)^3}\frac{m}{p^2} \label{gg0}
\end{equation}
where $g=4\pi a/m$ (we use $\hbar=k_{\rm B}=1$ everywhere). 

Let us denote by $\ket{\{n_{\pp\sigma}\}_{\pp\sigma}}_0$ the Fock eigenbasis of $\hat{H}_0$, that is,
the Fock states with particle occupation numbers $n_{\pp\sigma}$. Following Landau's argument \cite{Landau1956}, we view the quasiparticle 
states as perturbed particle states and expand them in powers of $\hat V$
\begin{equation}
\ket{\{n_{\pp\sigma}\}}= \ket{\{n_{\pp\sigma}\}}_0+\ket{\{n_{\pp\sigma}\}}_1+\ket{\{n_{\pp\sigma}\}}_2+\ldots
\end{equation}
where the memory of the original occupation numbers 
is retained throughout the adiabatic switching on of the interactions. 

\textit{Energy, interaction functions and collision amplitudes of the quasiparticles---} 
We first relate the Fermi liquid parameters to the matrix elements of $\hat H$
in the (orthonormal) basis of the quasiparticle states $\ket{\{n_{\pp\sigma}\}}$. To define the energy $\epsilon_{\pp\sigma}$
of the quasiparticles in our quantum formalism,
we consider the diagonal elements of $\hat H$
in the Fermi sea:
\be
\ket{{\rm FS}}=\ket{\{n_{\pp\sigma}=n_p^0\}}  \text{ with } n_{p}^0=\Theta(p_{\rm F}-p) 
\ee
and in the states $\ket{\pp\sigma}$ and $\ket{\overlinef{\pp\sigma}}$ with either one quasiparticle or one quasihole excitation in mode $\pp\sigma$
and the rest of the modes $\pp'\sigma'\neq\pp\sigma$ occupied as in the Fermi sea:
\bea
\ket{\pp\sigma}&=&\ket{n_{\pp\sigma}=1,\{n_{\pp'\sigma'}=n_{p'}^0\}_{\pp'\sigma'\neq\pp\sigma}} \\
\ket{\overlinef{\pp\sigma}}&=&\ket{n_{\pp\sigma}=0,\{n_{\pp'\sigma'}=n_{p'}^0\}_{\pp'\sigma'\neq\pp\sigma}}
 \label{psigma}
\eea
The eigenenergy $\epsilon_{\pp\sigma}^0$ is by definition the cost
of adding\footnote{
Our states $\ket{\pp\sigma}$ and $\ket{\overlinef{\pp\sigma}}$ are constructed with the Fermi sea as the reference state. More general choices
of the reference state $\ket{\psi}=\ket{\{n_{\pp'\sigma'}\}}$ are possible, in particular if one is interested in the quasiparticle energy $\epsilon_{\pp\sigma}$ away from the ground state.
In this case $\ket{\pp\sigma}\equiv(1-n_{\pp\sigma})\ket{\psi({\overline{\pp\sigma}}})+n_{\pp\sigma}\ket{\psi}$, where $n_{\pp\sigma}$ is the occupation of mode $\pp\sigma$ in $\ket{\psi}$ and $\ket{\psi({\overline{\pp\sigma}}})$ has the inverse occupation $1-n_{\pp\sigma}$ in $\pp\sigma$ and the same occupation $n_{\pp'\sigma'}$ as $\ket{\psi}$ in $\pp'\sigma'\neq\pp\sigma$.}
the quasiparticle $\pp\sigma$:
\be
\epsilon_{\pp\sigma}^0\equiv\bra{\pp\sigma}\hat H \ket{\pp\sigma}-
\bra{\overlinef{\pp\sigma}}\hat H \ket{\overlinef{\pp\sigma}}
\ee
Similarly, for the interaction function $f_{\sigma\sigma'}(\pp,\pp')$, we introduce the two-quasiparticle states $\ket{\pp\sigma,\pp'\sigma'}$, 
$\ket{\overlinef{\pp\sigma},\overlinef{\pp'\sigma'}}$, $\ket{\overlinef{\pp\sigma},\pp'\sigma'}$
and $\ket{\pp\sigma,\overlinef{\pp'\sigma'}}$ that coincide with the Fermi sea except for the occupation of modes $\pp\sigma$ and $\pp'\sigma'$.
We then view $f_{\sigma\sigma'}(\pp,\pp')$ as
the cost of adding quasiparticle $\pp\sigma$ in the presence of $\pp'\sigma'$ minus
the cost of adding it in absence of $\pp'\sigma'$:
\begin{widetext}
\be
L^3f_{\sigma\sigma'}(\pp,\pp')\equiv\bra{\pp\sigma,\pp'\sigma'}\hat H \ket{\pp\sigma,\pp'\sigma'}
-\bra{\overlinef{\pp\sigma},\pp'\sigma'}\hat H \ket{\overlinef{\pp\sigma},\pp'\sigma'}
-\bb{\bra{\pp\sigma,\overlinef{\pp'\sigma'}}\hat H \ket{\pp\sigma,\overlinef{\pp'\sigma'}}
-\bra{\overlinef{\pp\sigma},\overlinef{\pp'\sigma'}}\hat H \ket{\overlinef{\pp\sigma},\overlinef{\pp'\sigma'}}} \label{fpp}
\ee
\end{widetext}
We recognize in Eq.~\eqref{fpp} the second derivative $\partial^2 E/\partial n_{\pp\sigma}\partial n_{\pp'\sigma'}$ with respect to the discreet variables $ n_{\pp\sigma}$,  $n_{\pp'\sigma'}$ \cite{LandauLipschitzVol9}.

Even to high order in $\hat{V}$ the quasiparticles states $\ket{\{n_{\pp\sigma}\}}$
are never exact eigenstates of $\hat{H}$:
in an ergodic system, the eigenstates are exclusively 
characterized by their energy $E$ according to the eigenstate thermalization hypothesis, 
and should be viewed as superpositions of all the quasiparticle states at that energy.
The transitions ensuring ergodicity are described by the off-diagonal elements
of $\hat H$, which remain nonzero on the energy shell even in the quasiparticle basis. Unlike the particle states $\ket{\{n_{\pp\sigma}\}}_0$, the quasiparticle states $\ket{\{n_{\pp\sigma}\}}$ (at least the low-energy ones) are
long-lived: 
transitions between them are rare, occurring on timescales of the order of the collision time $\tau\propto1/T^2$ (see \eqqref{tau}), much longer
than the inverse eigenfrequencies $\approx1/T$ of the transitions between thermally occupied states. 
Thus, when the system is prepared in an initial quasiparticle state $\ket{\psi(0)}=\ket{i}$ of energy $E_i$, 
we can write its state at time $1/T\ll t \ll \tau$ as $\ket{\psi(t)}=\eee^{-\ii E_i t}\ket{i}+\ket{\delta\psi(t)}$ where $\ket{\delta\psi(t)}=O(t/\tau)$. Then the probability $w_{i\to f}$
of observing the system in a final state $\ket{f}\neq\ket{i}$ of energy $E_f$ obeys Fermi's golden rule:
\begin{equation}
\frac{\dd w_{i\to f}}{\dd t}=2\pi|\mathcal{A}_{i\to f}|^2\delta(E_i-E_f)\text{ where } \mathcal{A}_{i\to f}=\bra{f}\hat{H}\ket{i}
\end{equation}

\textit{Perturbative expansion of $f$ and $\mathcal{A}$---}
In a weakly-interacting Fermi gas, we compute the 
interaction function $f$ 
and transition amplitudes $\mathcal{A}_{i\to f}$ 
to second-order in $\hat{V}$. 
A generic quasiparticle state $\ket{i}$ can be written in the form\footnote{The second-order
corrections to the eigenstates affect the expression of $\bra{f}\hat H\ket{i}$ but can be eliminated 
using the orthonormality relation ${}_1\braket{f}{i}_1+{}_2\braket{f}{i}_0+{}_0\braket{f}{i}_2=0$.}
\begin{equation}
\ket{i}=\bbcro{1+\frac{\hat Q_i}{E_i^0-\hat H_0}\hat V}\ket{i}_0 +O(\hat V^2)
\end{equation}
where $E_i^0={}_0\bra{i}\hat H_0 \ket{i}_0$ and
$\hat Q_i$ projects orthogonally to degenerate subspace of $\ket{i}_0$. 

Using this expansion for initial and final states with $ E_i=E_f $ (and possibly $\ket{i}=\ket{f}$), we obtain:
\begin{equation}
\bra{f}\hat H\ket{i}={}_0\bra{f}\hat H_0+ \hat V+\hat V \hat Q_i \frac{1}{E_i^0-\hat H_0} \hat Q_f \hat V \ket{i}_0+O(\hat  V^3)  \label{Aif}
\end{equation}
where one can recognize the second-order expansion of the $T$-matrix,
or the effective Hamiltonian from Ref.~\cite{Cohen} (see Sec.~$\text{B}_{\rm I}$ therein).
The calculation of $f_{\sigma\sigma'}$ from here is standard\footnote{
The standard procedure to regularize the UV divergences stemming from the contact
potential is recalled in Appendix \ref{appicoll}} \cite{LandauLipschitzVol9}. For $p=p'=p_{\rm F}$ and $\pp\neq\pp'$,
we have:
\begin{eqnarray}
\frac{f_{\upa\dwa}(\pp,\pp')}{g}\!\!&=&\!\!1+\frac{2k_{\rm F} a}{\pi} \bbcro{I(\theta)+J(\theta)}+O(a^2)\label{fupdw}\\
\frac{f_{\sigma\sigma}(\pp,\pp')}{g}\!\!&=&\!\!\frac{2k_{\rm F} a}{\pi} J(\theta)+O(a^2) \label{fupup}
\end{eqnarray}
where $\theta=(\widehat{\pp,\pp'})$ and we introduced the functions $I$, $J$ 
\begin{eqnarray}
I(\theta)&=&1-\frac{\sin\theta/2}{2}\text{ln}\frac{1+\sin\theta/2}{1-\sin\theta/2}\\
J(\theta)&=&\frac{1}{2}\bb{1+\frac{\cos^2\theta/2}{2\sin\theta/2}\text{ln}\frac{1+\sin\theta/2}{1-\sin\theta/2}}
\end{eqnarray}

We now turn to the transition amplitudes, described by \eqqref{Aif} for $\ket{i}\cancel{\propto}
\ket{f}$. The only relevant transitions  at low temperatures are $2\leftrightarrow2$ transitions where quasiparticles of momenta and spin $\pp_3\sigma_3,\pp_4\sigma_4$ scatter into modes $\pp_1\sigma_1,\pp_2\sigma_2$. Since the total spin is conserved, we can impose that $\sigma_1=\sigma_4$ and $\sigma_2=\sigma_3$ without loss of generality. We choose as initial and final states
\begin{eqnarray}
\!\!\!\!\!\!\!\!\!\!\!\ket{i_{2\leftrightarrow2}}_0\!\!&=&\!\!\ket{\overlinef{\pp_1\sigma_1},\overlinef{\pp_2\sigma_2},\pp_3\sigma_2,\pp_4\sigma_1}_0\\
\!\!\!\!\!\!\!\!\!\!\!\ket{f_{2\leftrightarrow2}}_0\!\!&=&\!\!\hat a_{\pp_1\sigma_1}^\dagger \hat a_{\pp_2\sigma_2}^\dagger \hat a_{\pp_3\sigma_2} \hat a_{\pp_4\sigma_1}\ket{i_{2\leftrightarrow2}}_0
\end{eqnarray}
This choice of $\ket{i}$ and $\ket{f}$ corresponds to the collision
amplitude in the ground state $\ket{\text{FS}}$,
which is sufficient for the low-temperature transport properties.

\begin{figure}
\begin{tikzpicture}[xscale=1.3,yscale=1.5]
\coordinate (A) at (-0.5,0);
\coordinate (B) at (0.5,0);
\def\xA{-0.5}
\def\yA{0}
\def\xB{0.5}
\def\yB{0}
\draw (A) node{\textbullet};
\draw (B) node{\textbullet};
\draw[->-] (A) arc (135:45:{1/sqrt(2.)}) ;
\draw (0,0.4) node {$\pp_a\!\!\upa$};
\draw (0,-0.4) node {$\pp_b\!\!\dwa$};
\draw[->-] (A) arc (-135:-45:{1/sqrt(2.)});
\draw (\xA-0.75,\yA+0.5) node {I};
\draw[->-] (\xA-0.5,\yA+0.5) -- node[above=0.1cm]{$\pp_{3}\!\!\dwa$} (A);
\draw[->-] (\xA-0.5,\yA-0.5) -- node[below=0.1cm]{$\pp_{4}\!\!\upa$} (A);
\draw[->-] (B) -- node[above=0.1cm]{$\pp_{1}\!\!\upa$} (\xB+0.5,\yB+0.5);
\draw[->-] (B) -- node[below=0.1cm]{$\pp_{2}\!\!\dwa$} (\xB+0.5,\yB-0.5);
\draw (0,-1) node { {\scriptsize $\ket{n_{\rm I}}_0\!=\!\hat a_{\pp_a\upa}^\dagger \hat a_{\pp_b\dwa}^\dagger \hat a_{\pp_3\dwa} \hat a_{\pp_4\upa}\ket{i}_0$}};
\end{tikzpicture}
\begin{tikzpicture}[xscale=1.3,yscale=1.5]
\coordinate (A) at (1,0);
\coordinate (B) at (-1,0);
\def\xA{-0.5}
\def\yA{0}
\def\xB{0.5}
\def\yB{0}
\draw (A) node{\textbullet};
\draw (B) node{\textbullet};
\draw[->-] (B) arc (120:60:{2}) ;
\draw (0,0.4) node {$\pp_a\!\!\upa$};
\draw (0,-0.4) node {$\pp_b\!\!\dwa$};
\draw (\xB+0.75,\yA+0.5) node {II};
\draw[->-] (B) arc (-120:-60:{2});
\draw[->-] (\xA+0.7,\yA+0.5) -- node[above=0.1cm]{$\pp_{3}\!\!\dwa$} (A);
\draw[->-] (\xA+0.7,\yA-0.5) -- node[below=0.1cm]{$\pp_{4}\!\!\upa$} (A);
\draw[->-] (B) -- node[above=0.1cm]{$\pp_{1}\!\!\upa$} (\xB-0.7,\yB+0.5);
\draw[->-] (B) -- node[below=0.1cm]{$\pp_{2}\!\!\dwa$} (\xB-0.7,\yB-0.5);
\draw (0,-1) node {\scriptsize $\ket{n_{\rm II}}_0\!=\!\hat a_{\pp_1\upa}^\dagger \hat a_{\pp_2\dwa}^\dagger \hat a_{\pp_b\dwa} \hat a_{\pp_a\upa}\ket{i}_0$};
\end{tikzpicture}

\begin{tabular}[b]{cc}
\hspace{-0.3cm}
\begin{tikzpicture}[xscale=1.3,yscale=1.1]
\coordinate (A) at (-0.5,0.5);
\coordinate (B) at (0.5,-0.5);
\def\xA{-0.5}
\def\yA{0.5}
\def\xB{0.5}
\def\yB{-0.5}
\draw (A) node{\textbullet};
\draw (B) node{\textbullet};
\draw (0.6,0.4) node {$\pp_a,-\sigma_2$};
\draw (-0.5,-0.4) node {$\pp_b,-\sigma_1$};
\draw[-<-] (A) arc (90:0:{1});
\draw[->-] (A) arc (-180:-90:{1});
\draw[->-] (\xA-0.5,\yA+0.5) -- node[left=0.1cm]{$\pp_{3}\sigma_2$} (A);
\draw[->-] (\xB-0.5,\yB-0.5) -- node[below=0.1cm]{$\pp_{4}\sigma_1$} (B);
\draw[->-] (A) -- node[above=0.1cm]{$\pp_{1}\sigma_1$} (\xA+0.5,\yA+0.5);
\draw[->-] (B) -- node[right=0.1cm]{$\pp_{2}\sigma_2$} (\xB+0.5,\yB-0.5);
\draw (-1.1,-1) node {III};
\draw (-0.4,-1.7)  node {\scriptsize $\ket{n_{\rm III}}_0=\hat a_{\pp_1\sigma_1}^\dagger \hat a_{\pp_b,-\sigma_1}^\dagger \hat a_{\pp_3\sigma_2} \hat a_{\pp_a,-\sigma_2}\ket{i}_0$};
\draw (-0.4,-2.5)  node {};
\end{tikzpicture}
& \hspace{-2cm}
\begin{tikzpicture}[xscale=1.3,yscale=1.1]
\coordinate (A) at (0.5,0.5);
\coordinate (B) at (-0.5,-0.5);
\def\xA{0.5}
\def\yA{0.5}
\def\xB{-0.5}
\def\yB{-0.5}
\draw (A) node{\textbullet};
\draw (B) node{\textbullet};
\draw (0.7,-0.3) node {$\pp_a,-\sigma_2$};
\draw (-0.6,0.3) node {$\pp_b,-\sigma_1$};
\draw[-<-] (A) arc (0:-90:{1});
\draw[->-] (A) arc (90:180:{1});
\draw[->-] (\xA-0.5,\yA+0.5) -- node[left=0.1cm]{$\pp_{3}\sigma_2$} (A);
\draw[->-] (\xB-0.5,\yB-0.5) -- node[below=0.1cm]{$\pp_{4}\sigma_1$} (B);
\draw[->-] (A) -- node[above=0.1cm]{$\pp_{1}\sigma_1$} (\xA+0.5,\yA+0.5);
\draw[->-] (B) -- node[right=0.1cm]{$\pp_{2}\sigma_2$} (\xB+0.5,\yB-0.5);
\draw (1,-1) node {IV};
\draw (-0.4,-2.3)  node {\scriptsize $\ket{n_{\rm IV}}_0=\hat a_{\pp_a,-\sigma_2}^\dagger \hat a_{\pp_2\sigma_2}^\dagger \hat a_{\pp_b,-\sigma_1} \hat a_{\pp_4\sigma_1}\ket{i}_0$};
\end{tikzpicture}
\end{tabular}

\begin{tabular}[b]{cc}
\hspace{-0.3cm}
\begin{tikzpicture}[xscale=1.3,yscale=1.1]
\coordinate (A) at (-0.5,0.5);
\coordinate (B) at (0.5,-0.5);
\def\xA{-0.5}
\def\yA{0.5}
\def\xB{0.5}
\def\yB{-0.5}
\draw (A) node{\textbullet};
\draw (B) node{\textbullet};
\draw (0.4,0.4) node {$\pp_a\dwa$};
\draw (-0.4,-0.4) node {$\pp_b\dwa$};
\draw[-<-] (A) arc (90:0:{1});
\draw[->-] (A) arc (-180:-90:{1});
\draw[->-] (\xA-0.5,\yA+0.5) -- node[left=0.1cm]{$\pp_{4}\upa$} (A);
\draw[->-] (\xB-0.5,\yB-0.5) -- node[below=0.1cm]{$\pp_{3}\upa$} (B);
\draw[->-] (A) -- node[above=0.1cm]{$\pp_{1}\upa$} (\xA+0.5,\yA+0.5);
\draw[->-] (B) -- node[right=0.1cm]{$\pp_{2}\upa$} (\xB+0.5,\yB-0.5);
\draw (-1.1,-1) node {V};
\draw (-0.4,-1.5)  node {\scriptsize $\ket{n_{\rm V}}_0=\hat a_{\pp_1\upa}^\dagger \hat a_{\pp_b\dwa}^\dagger \hat a_{\pp_4\upa} \hat a_{\pp_a\dwa}\ket{i}_0$};
\end{tikzpicture}
& \hspace{-0.4cm}
\begin{tikzpicture}[xscale=1.3,yscale=1.1]
\coordinate (A) at (0.5,0.5);
\coordinate (B) at (-0.5,-0.5);
\def\xA{0.5}
\def\yA{0.5}
\def\xB{-0.5}
\def\yB{-0.5}
\draw (A) node{\textbullet};
\draw (B) node{\textbullet};
\draw (0.5,-0.3) node {$\pp_a\dwa$};
\draw (-0.5,0.3) node {$\pp_b\dwa$};
\draw[-<-] (A) arc (0:-90:{1});
\draw[->-] (A) arc (90:180:{1});
\draw[->-] (\xA-0.5,\yA+0.5) -- node[left=0.1cm]{$\pp_{4}\upa$} (A);
\draw[->-] (\xB-0.5,\yB-0.5) -- node[below=0.1cm]{$\pp_{3}\upa$} (B);
\draw[->-] (A) -- node[above=0.1cm]{$\pp_{1}\upa$} (\xA+0.5,\yA+0.5);
\draw[->-] (B) -- node[right=0.1cm]{$\pp_{2}\upa$} (\xB+0.5,\yB-0.5);
\draw (1,-1) node {VI};
\draw (0.4,-1.5)  node {\scriptsize $\ket{n_{\rm VI}}_0=\hat a_{\pp_a\dwa}^\dagger \hat a_{\pp_2\upa}^\dagger \hat a_{\pp_b\dwa} \hat a_{\pp_3\upa}\ket{i}_0$};
\end{tikzpicture}
\end{tabular}
\caption{\label{diagrammes} Diagrams representing the four intermediate states of the $\pp_4\sigma_1,\pp_3\sigma_2\to\pp_1\sigma_1,\pp_2\sigma_2$ transition to second-order in $\hat V$.}
\end{figure}
We compute the corresponding transition amplitude 
\begin{equation}
\frac{1}{L^3}\mathcal{A}_{\sigma_1\sigma_2}(\pp_1,\pp_2|\pp_3,\pp_4)\equiv\mathcal{A}_{\ket{i_{2\leftrightarrow2}}\to \ket{f_{2\leftrightarrow2}}}
\end{equation}
by inserting in Eq.~\eqref{Aif}
the closure relation on the particle Fock states $\sum_{\ket{n}_0} \ket{n}_0\bra{n}_0=\text{Id}$.
For collisions between quasiparticles of opposite spin we have $\mathcal{A}_{\upa\dwa}=\mathcal{A}_{\dwa\upa}$ by spin-symmetry and
there are four possible intermediate states $\ket{n_i},\ i=\textrm{I,\ II,\ III,\ IV}$  depicted diagrammatically and written explicitly on Fig.~\ref{diagrammes}.
This leads to
\begin{equation}
\!\!\!\frac{\mathcal{A}_{\upa\dwa}(\pp_1,\pp_2|\pp_3,\pp_4)}{g}=1+\frac{2k_{\rm F}a}{\pi}\bbcro{I(\theta_{12})+J(\theta_{13})}+O(a^2) \label{Aupdw}
\end{equation}
where we introduced the angles $\theta_{ij}=(\widehat{\pp_i,\pp_j})$.

As $\hat V$ couples only opposite-spin particles, quasiparticle transitions involving identical spins arise solely to second order in
$\hat V$ and, in an unpolarized system, we have
$\mathcal{A}_{\upa\upa}=\mathcal{A}_{\dwa\dwa}$. Diagrams I and II
are impossible (they would involve the simultaneous annihilation of two particles of same spin) 
and there appears instead diagrams V and VI which follow from III and IV by swapping momenta $\pp_3$ and $\pp_4$, which now play symmetric roles.
This results in
\begin{multline}
\!\!\!\frac{\mathcal{A}_{\upa\upa}(\pp_1,\pp_2|\pp_3,\pp_4)}{g}=\frac{2k_{\rm F}a}{\pi}\bbcro{J(\theta_{13})-J(\theta_{14})}+O(a^3) \label{Aupup}
\end{multline}
This amplitude is antisymmetric with respect to the exchange of $\pp_1$ and $\pp_2$ or
$\pp_3$ and $\pp_4$, as expected for the scattering of indistinguishable
fermions. 

The amplitudes \eqqrefs{Aupdw}{Aupup} are often approximated by their forward scattering limit,
that is their finite and nonzero value when $(\pp_1,\pp_2)\to(\pp_4,\pp_3)$:
\begin{eqnarray}
\!\!\!\!\!\frac{\mathcal{A}_{\upa\dwa}(\pp,\pp'|\pp',\pp)}{g}\!\!&=&\!\!1+\frac{2k_{\rm F}a}{\pi}\bbcro{I(\theta)+J(\theta)}+O(a^2)\label{Aupdwforward}\\
\!\!\!\!\!\frac{\mathcal{A}_{\upa\upa}(\pp,\pp'|\pp',\pp)}{g}\!\!&=&\!\!\frac{2k_{\rm F}a}{\pi}\bbcro{J(\theta)-{1}}+O(a^2) \label{Aupupforward}
\end{eqnarray}
Remarkably, the forward scattering value of $\mathcal{A}_{\upa\upa}$ does not coincide with 
$f_{\upa\upa}$. This is due to the absence of diagrams V and VI in the calculation
of diagonal matrix elements appearing in the expression \eqref{fpp} of $f_{\upa\upa}$: the intermediate states $\ket{n_{\rm V}}$ and
$\ket{n_{\rm VI}}$ become proportional to the initial state $\ket{i}$, and are thus excluded from the perturbative treatment.
A similar issue arises for $\mathcal{A}_{\upa\dwa}$ but only to higher order in $g$.
Landau derived a Bethe-Salpeter equation (see Eq.~(33) in \cite{Landau1959} and the End Matter) that performs a resummation of the missing diagrams,
thereby recovering the forward-scattering
value of $\mathcal{A}_{\sigma\sigma'}$ 
from the knowledge of $f_{\sigma\sigma'}$ only.
Despite its elegance, this approach does not allow for a controlled calculation such as ours
 since the general expressions (\ref{Aupdw}--\ref{Aupup}) are needed for a quantitative solution
of the transport equation.

\textit{Navier-Stokes equations of the Fermi liquid}---
We now turn to the transport dynamics. From an initial
thermal state, described by the density matrix $\hat\rho_{\rm eq}$,
the system is brought out-of-equilibrium $\hat\rho(t)=\hat\rho_{\rm eq}+\delta\hat\rho(t)$ 
by a weak external field $U_\sigma(\rr,t)$ of frequency $\omega$.
We introduce the quasiparticle distribution function:
\begin{equation}
n_\sigma(\pp,\qq,t)=\text{Tr}\bb{\hat\rho(t)\hat\gamma_{\pp-\frac{\qq}{2}\sigma}^\dagger\hat\gamma_{\pp+\frac{\qq}{2}\sigma}}
\end{equation}
where $\hat \gamma_{\pp\sigma}$ annihilates a quasiparticle in mode $\pp\sigma$. At long wavelengths ($p_{\rm F}q/m^*\ll T$ with $q$  the wavenumber of the perturbation), $n_\sigma$ obeys the linearized transport equation \cite{Sauls2022,Nishida2021}
\begin{multline}
\bb{\omega -\frac{\pp\cdot\qq}{m^*}}n_\sigma(\pp)+\frac{\partial n_{\rm eq}}{\partial \epsilon}\Big\vert_{\epsilon=\epsilon_{\pp\sigma}^0}\frac{\pp\cdot\qq}{m^*}\Bigg[U_{\sigma}(\qq)\\
+\frac{1}{L^3}\sum_{\pp'\sigma'}f_{\sigma\sigma'}(\pp,\pp')n_{\sigma'}(\pp')\Bigg]=\ii I_{\sigma}(\pp) \label{transport}
\end{multline}
where $n_{\rm eq}(\epsilon)=1/(1+\eee^{(\epsilon-\mu)/T})$ is the Fermi-Dirac distribution and the driving field is Fourier-transformed according to
$U_\sigma(\rr,t)=\sum_{\qq} U_\sigma(\qq) \eee^{\ii(\qq\cdot\rr-\omega t)}$. We have used the fact that $n_\sigma(\pp,\qq,t)$ is proportional to $\eee^{-\ii\omega t}$ as a consequence of the periodic driving.
The effective mass $m^*$ such that $(\partial\epsilon_{\pp\sigma}^0/\partial p)_{p=p_{\rm F}}=p_{\rm F}/m^*$
is related to $m$ by $m^\ast/m=\bar F_1^+$, where $\bar F_l^\pm=1+F_l^\pm/(2l+1)$ and
\begin{equation}
F_l^\pm=\frac{m^*p_{\rm F}}{\pi^2}\int_0^\pi\sin\theta\dd\theta\frac{P_l(\cos\theta)}{||P_l||^2}\frac{f_{\upa\upa}(\theta)\pm f_{\upa\dwa}(\theta)}{2}
\end{equation}
are the Landau parameters, \textit{i.e.} the decomposition of $f_{\upa\upa}\pm f_{\upa\dwa}$ onto the Legendre polynomials $P_l$.
The collision integral $I_{\sigma}(\pp)$ (given explicitly in the End Matter) describes on shell $2\leftrightarrow2$ transitions
with transition probabilities 
\be
W_{\sigma\sigma'}=|\mathcal{A}_{\sigma\sigma'}|^2
\ee
Its magnitude compared to
the other terms of the transport equation is controlled by the mean collision time
\begin{equation}
\frac{1}{\tau}=\frac{(m^*)^3T^2}{(2\pi)^3}\int_0^\pi \int_0^{2\pi} \frac{\sin\theta\dd\theta\dd\phi}{4\pi} \frac{w_+(\theta,\phi)}{2\cos(\theta/2)} \label{tau}
\end{equation}
In the symmetrized collision probability $w_+$,
the contribution of $W_{\upa\upa}=O(a^4)$ can be neglected 
even compared to the subleading corrections $O(a^3)$ to $W_{\upa\dwa}$,
such that
\begin{multline}
w_+(\theta_{12},\phi_{12})=\frac{1}{2}\Big[W_{\upa\dwa}(\pp_1\pp_2|\pp_3\pp_4)+ W_{\upa\dwa}(\pp_1\pp_2|\pp_4\pp_3)\Big] \label{wplus}
\end{multline}
in which $\phi_{12}=(\widehat{\pp_1-\pp_2,\pp_3-\pp_4})$ is an azimuthal angle.
The forward-scattering limit $\mathcal{A}_{\upa\dwa}(\pp_1,\pp_2|\pp_3,\pp_4)\approx \mathcal{A}_{\upa\dwa}(\pp_1,\pp_2|\pp_2,\pp_1)$
amounts to neglecting the dependence
of the collision probability on this azimuthal angle $w_+(\theta,\phi)\approx w_+(\theta,0)$;
from \eqqref{tau}, it is now clear that this constitutes an uncontrolled approximation
on the mean collision time. The error is greater on transport coefficients,
where the azimuthal dependence matters even more.

The exact solution of \eqqref{transport} can be obtained
by decomposing the quasiparticle distribution onto a basis of 
orthogonal polynomials \cite{Hofmann2023,Hofmann2025,theserepplinger,disphydro}. In Ref.~\cite{disphydro}
this exact decomposition is performed for the total density $n_\upa+n_\dwa$ to leading order in $g$, hence for
isotropic interaction functions $f_{\sigma\sigma'}$ and collision probability $W_{\upa\dwa}$. We generalize this
method to arbitrary $f_{\sigma\sigma'}$ and $W_{\upa\dwa}$ and to the polarisation $n_\upa-n_\dwa$ in the End Matter.
In the hydrodynamic limit, $\omega_0\tau\ll1$ where $\omega_0=p_{\rm F}q/m^*$ is a typical 
excitation frequency, collisions dominate and prevent all non-conserved quantities
from being significantly nonzero. For unidimensional longitudinal flows, the dynamics of the whole distribution $n_\sigma$
reduces to a set of hydrodynamic equations on the four conserved
quantities: the total density $\rho_+=(1/V)\sum_{\pp\sigma} n_\sigma(\pp)$,
the polarisation $\rho_-=(1/V)\sum_\pp [n_\upa(\pp)- n_\dwa(\pp)]$, the longitudinal velocity 
$v_\parallel=({1}/{V})\sum_{\pp\sigma} (\pp\cdot\qq/m\rho_0 q) n_\sigma(\pp)$
and the energy density $e=(1/V)\sum_{\pp\sigma} (\frac{p^2}{2m^*}-\mu) n_\sigma(\pp)$.
We obtain the Fermi-liquid equivalents of the linearized Navier-Stokes equations:
\bea
\frac{\partial \rho_+}{\partial t}+\rho_0\frac{\partial v_{\parallel}}{\partial z}&=&0 \label{NS1} \\
m\rho_0\frac{\partial v_{\parallel}}{\partial t}+\frac{1}{\rho_0\chi}\frac{\partial \rho_+}{\partial z}-\sum_\sigma \rho_{\sigma,0}\frac{\partial U_\sigma}{\partial z}&=&\frac{4}{3}\eta\frac{\partial^2 v_{\parallel}}{\partial z^2} \label{NS2}\\
c_V\frac{\partial e}{\partial t}&=& {\kappa} \frac{\partial^2 e}{\partial z^2} \label{NS3}
\eea
\be
\frac{\partial \rho_-}{\partial t} = D\frac{\partial^2\rho_-}{\partial z^2} -\frac{ D}{2}\chi_p\frac{\partial^2 U_-}{\partial z^2} \label{NS4}
\ee
where $U_\pm=U_\upa\pm U_\dwa$.
The leading terms in the Chapman-Enskog expansion in powers of $\omega_0\tau$
are gathered on the left-hand side of Eqs.~(\ref{NS1}--\ref{NS4}); they involve
static parameters such the specific heat $c_V=m^*\pF T/3$
or compressibility $\chi^{-1}\equiv\rho_0(\dd \mu/\dd \rho)=\pF^2\bar F_0^+/3mm^\ast$.
Conversely, the dissipative
terms on the right-hand side of Eqs.~(\ref{NS1}--\ref{NS4}) 
stem from an inversion of the collision kernel and are therefore subleading
in $\omega_0\tau$. The three transport coefficients that control them are
given in Table \ref{table}. Note that the contribution of the drive $U_-$ to the spin diffusion \eqqref{NS4}
also involves the polarisation suceptibility $\chi_p=\left[{\dd (\rho_{{\rm eq},\upa}-\rho_{{\rm eq},\dwa})}/{\dd (\mu_\upa-\mu_\dwa)}\right]_{\mu_\upa=\mu_\dwa}$. Corrections 
of order $(\omega_0\tau)^2$ responsible for Burnett's 
hydrodynamics \cite{disphydro,FochFord} are omitted from Eqs.~(\ref{NS1}--\ref{NS4}).

The second column of Tab.~\ref{table} gives a non-perturbative expression of $\eta$, $\kappa$, $D$
in terms of collision times $\tau_\eta$, $\tau_\kappa$ and $\tau_D$.
These collision time are all comparable to $\tau$, and
their expression in terms of the collision kernel (that is ultimately in terms of $W_{\upa\dwa}$)
is given by Tab.~\ref{tabannexe} in the End Matter. Importantly, our exact expressions of $\eta$ and $D$
\textit{disagree} with the classical result of Refs.~\cite{Sykes1968,Wilkins1968}
in which the contribution of the quantum force $(1/V)\sum_{\pp'\sigma'}f_{\sigma\sigma'}(\pp,\pp')n_{\sigma'}(\pp')$
to dissipation is omitted, such that the factors $\bar F_2^+$ and $\bar F_1^-$ are missing
in $\eta$ and $D$ respectively. This corrections illustrates the reliability of our systematic method for solving
the transport equation.

\onecolumngrid
\vspace{\columnsep}
\begin{center}
\begin{table}[h!]
\begin{tabular}{c|c|@{\hspace{12pt}}c@{\hspace{12pt}}|c}
\hline
\hline
{Transport coefficient}  & {Exact  expression} & \makecell{Perturbative value \\ ($\bar a=\kF a$, $\tau_\sigma=1/8ma^2T^2$)} & {Conserved  quantity} \\
\hline
&&&\\[-0.2cm]
{$\eta$}  & {$ \frac{\bar{F}_2^+}{5}\frac{p_{\rm F}^2\tau_\eta }{m^*}$} & { $\!\!\!0.5153[1\!+\!0.2633\bar a\!+\!O(a^2)]\frac{p_{\rm F}^2\tau_\sigma}{2m}\rho_0$} & {$v_{\parallel},\rho_+$}  \\
&&&\\[-0.2cm]
{$\kappa$}   & {$ \frac{p_{\rm F}^3\tau_\kappa}{9m^\ast}$} & { $\!\! 0.0831\bbcro{1\!-\!0.0875\bar a\!+\!O(a^2)}\frac{p_{\rm F}^3 \tau_\sigma }{m}T$} & {$e$}  \\
&&&\\[-0.2cm]
{$D$} & {$\frac{\bar F^-_0 \bar F^-_1 {p_{\rm F}^2}\tau_D}{3(m^\ast)^2}$} & { $0.268\!\bbcro{1\!-\!0.7848 \bar a\!+\!O(a^2)}\!\frac{p_{\rm F}^2 \tau_\sigma}{m^2}$} & {$\rho_-$}   \\
\hline
\hline
\end{tabular}
\caption{Exact and perturbative expression of the shear viscosity (top row), thermal conductivity (middle row) and spin diffusivity (bottom row).
\label{table}}
\end{table}
\end{center}
\vspace{\columnsep}
\twocolumngrid

Then, the third column of Tab.~\ref{table} gives the perturbative expansion
of the transport coefficients. This is obtained by injecting
the perturbative results \eqref{fupdw}, \eqref{fupup} and \eqref{Aupdw}
in the expression of the Landau parameters $F_l^\pm$
and collision times $\tau_\eta$, $\tau_\kappa$ and $\tau_D$ (see Tab.~\ref{tabannexe}).
Our perturbative expression for $\kappa$ agrees with 
Ref.~\cite{Mohling1976} (Eq.~(A14) therein) while for $\eta$ we neatly disagree\footnote{We have identified an error in Eq.~(A12)
of \cite{Mohling1976} (the 111/32 should be a 105/32) which affects the final value of the shear viscosity.}.
Since Ref.~\cite{Mohling1976} did not consider the spin dynamics,
there was (to our knowledge) no equivalent of our perturbative expression of $D$ in the literature.

Importantly, the corrections proportional to $k_{\rm F}a$ have non trivial signs, 
such that a given transport coefficient
may vary with $\kF a$ oppositely to the mean collision rate
\begin{equation}
\frac{\tau_\sigma}{\tau}\simeq\frac{1}{4\pi}\bbcro{1+0.574k_{\rm F} a+O(a^2)}
\end{equation}
where $\tau_\sigma=\pi/2m\sigma T^2$ and $\sigma=4\pi a^2$
is the scattering crosssection to leading order in $a$.

We finally compare our value of $\eta$
to the measurement by the Yale group
\cite{yaleexp}. The experimental value $(2/12\pi)(\eta/\rho)(2m/p_{\rm F}^2\tau_\sigma)=0.07(2)_{\rm stat}(3)_{\rm syst}$ (see Fig.~4b)
is obtained by extrapolating to $T=0$ measurements of the sound
attenuation down to $T/\TF=0.17$ at $k_{\rm F}a=-0.67$.
Our formula gives $0.071$ (compared to $0.086$ to leading order in $a$).
Although additional measurements at various values of $k_{\rm F}a$
are needed to conclude, our results seem to describe reasonably well the weakly-interacting side of the phase
diagram. This is consistent with the estimate of a convergence radius between $\kF |a|=0.5$
and $\kF |a|=1$ for the perturbative serie of the equation of state \cite{Schwenk2021}. 

There are, to the best of our knowledge, no data on $D$ in the Fermi liquid regime to compare our formula with,
but measurements of the (transverse) spin diffusivity are state-of-the art \cite{Thywissen2014,Thywissen2015},
for example by monitoring the relaxation of an imprinted polarisation pattern \cite{Kohl2013}, 
or the collision of fully-polarized clouds \cite{Zwierlein2011Universal}.
Conversely, existing measurements of $\kappa$ \cite{Thomas2022} rely on the coupling
between total-density $\rho_+$ and energy-density $e$ fluctuations which disappears
in the Fermi liquid regime due to the symmetry about the Fermi surface. A measurement protocol for $\kappa$
at low temperature is thus yet to be found.

Several extensions of our work to the 
collisionless regime \cite{Pethick1969}, to polarized Fermi liquids 
\cite{Grimm2008,Zwierlein2011}, to lower dimensions \cite{Randeria1992,Bruun2012,Moritz2020} 
or to the vicinity of the phase transition \cite{Emery1976,Thomas2015,Sauls2022}
appear relevant for experiments in quantum gases, nuclear or condensed matter.
Our characterisation of the BCS side of the BCS-BEC crossover
may also help locate lower bounds on the transport coefficients 
\cite{Thywissen2018}.

\begin{acknowledgements}
Fruitful discussions with Nicolas Dupuis, Ludovic Pricoupenko and Félix Werner are gratefully acknowledged.
H.K. acknowledges support from the French Agence Nationale de la Recherche (ANR), under grant ANR-23-ERCS-0005 (project DYFERCO)
\end{acknowledgements}

\appendix
\section{Exact solution of the transport equation}
\label{appicoll}

In this section, we generalize to non isotropic $f_{\sigma\sigma'}$ and $W_{\upa\dwa}$ the method of Ref.~\cite{disphydro}
for inverting the collision kernel.

\textit{Collision integral}---
Assuming $W_{\sigma\sigma}=0$, the collision integral in \eqqref{transport} is given by:
\begin{widetext}
\be
    I_{\sigma}(\pp) = -\Gamma(\pp)n_\sigma(\pp)-\frac{1}{L^3}\sum_{\pp'}\bbcro{E(\pp',\pp)n_{-\sigma}(\pp')-\sum_{\sigma'}S_{\sigma',\sigma}(\pp',\pp)n_{\sigma'}(\pp')}
\ee
This expression involves three different collision kernels
\begin{eqnarray}
    E(\pp,\pp') &=& \frac{2\pi}{L^3}\sum_{\pp_3,\pp_4}W_{\upa\dwa}(\pp,\pp'|\pp_3,\pp_4)\delta^{\pp+\pp'}_{\pp_3+\pp_4}\delta(\epsilon_\pp^0+\epsilon_{\pp'}^0-\epsilon^0_{\pp_3}-\epsilon^{0}_{\pp_4})\bbcro{n^{\rm eq}_{\pp'}\overline{n}^{\rm eq}_{\pp_3}\overline{n}^{\rm eq}_{\pp_4}+\overline{n}^{\rm eq}_{\pp'}n^{\rm  eq}_{\pp_3}n^{\rm eq }_{\pp_4}} \\
    S_{\upa\upa}(\pp,\pp')=S_{\dwa\dwa}(\pp,\pp')&=& \frac{2\pi}{L^3}\sum_{\pp_2,\pp_4}W_{\upa\dwa}(\pp,\pp_2|\pp_4,\pp')\delta^{\pp+\pp_2}_{\pp'+\pp_4}\delta(\epsilon_\pp^0+\epsilon_{\pp_2}^0-\epsilon^0_{\pp'}-\epsilon^{0}_{\pp_4})\bbcro{n^{\rm eq}_{\pp_2}\overline{n}^{\rm eq}_{\pp'}\overline{n}^{\rm eq}_{\pp_4}+\overline{n}^{\rm eq}_{\pp_2}n^{\rm eq}_{\pp'}n^{\rm eq }_{\pp_4}} \label{Supup} \\
        S_{\upa\dwa}(\pp,\pp')=S_{\dwa\upa}(\pp,\pp')&=& \frac{2\pi}{L^3}\sum_{\pp_2,\pp_4}W_{\upa\dwa}(\pp,\pp_2|\pp',\pp_4)\delta^{\pp+\pp_2}_{\pp'+\pp_4}\delta(\epsilon_\pp^0+\epsilon_{\pp_2}^0-\epsilon^0_{\pp'}-\epsilon^{0}_{\pp_4})\bbcro{n^{\rm eq}_{\pp_2}\overline{n}^{\rm eq}_{\pp'}\overline{n}^{\rm eq}_{\pp_4}+\overline{n}^{\rm eq}_{\pp_2}n^{\rm eq}_{\pp'}n^{\rm eq }_{\pp_4}} \label{Supdw}
\end{eqnarray}
\end{widetext}
with $n^{\rm eq}_\pp=n_{\rm eq}(\epsilon_\pp^0)$ and $\bar n=1-n$. The quasiparticule lifetime $\Gamma$ is related to $E$ by $\Gamma(\pp)=\sum_{\pp'}E(\pp,\pp')/L^3$.
In  $E$, the momenta $\pp_3$ and $\pp_4$ play symmetric roles, such that $W_{\upa\dwa}$ can be directly replaced
by the symmetrized probability $w_+$, \eqqref{wplus}. For $S$ however, we need to combine \eqqrefs{Supup}{Supdw} to form $S_\pm=( S_{\upa\dwa}\pm S_{\upa\upa})/2$
in which $W_{\upa\dwa}$ is replaced by
\begin{multline}
w_\pm(\theta_{12},\phi_{12})=\frac{1}{2}\Big[W_{\upa\dwa}(\pp_1\pp_2|\pp_3\pp_4)\pm W_{\upa\dwa}(\pp_1\pp_2|\pp_4\pp_3)\Big]
\end{multline}
The calculation of $E$ and $S_\pm$ at low temperature proceeds as in Ref.~\cite{theserepplinger} and results
in the decoupling of the energy and angular dependencies:
\begin{eqnarray}
    E(\pp,\pp') &=& \frac{T(m^*)^2}{2\pi p_{\rm F}}\mathcal{S}(\varepsilon,-\varepsilon')\Omega_{E}(\gamma) \\
    S_\pm(\pp,\pp')&=& \frac{T(m^*)^2}{2\pi p_{\rm F}}\mathcal{S}(\varepsilon,\varepsilon')\Omega_{S\pm}(\gamma)
\end{eqnarray}
Here, 
\be
\mathcal{S}(\varepsilon,\varepsilon') = \frac{\varepsilon-\varepsilon'}{2}\frac{\cosh \frac{\varepsilon}{2}}{\cosh \frac{\varepsilon'}{2} \sinh \frac{\varepsilon-\varepsilon'}{2}}
\ee 
depends only on the reduced energies $\varepsilon=(\epsilon_\pp^0-\mu)/T$ and $\varepsilon'=(\epsilon_{\pp'}^0-\mu)/T$, and the angle $\gamma=(\widehat{\pp,\pp'})$ enters the functions
\begin{eqnarray}
\Omega_{E}(\theta) \!\!&=&\!\! \int^{2\pi}_0 \frac{\textrm{d}\phi}{2\pi}\frac{w_+(\theta,\phi)}{2|\text{cos}{\frac{\theta}{2}}|} \label{OmegaEtheta}\\
\Omega_{S\pm}(\theta_{13}) \!\!&=&\!\! \int^{2\pi}_0 \frac{\textrm{d}\phi_{13}}{2\pi}\frac{w_\pm\Big(\theta(\theta_{13},\phi_{13}),\phi(\theta_{13},\phi_{13})\Big)}{2|\text{sin}{\frac{\theta_{13}}{2}}|} \label{OmegaStheta}
\end{eqnarray}
Since $\pp$ and $\pp'$ play the role of $\pp_1$ and $\pp_3$ in \eqqref{Supdw},
the function $\Omega_{S\pm}$ is expressed in terms
of the angles $\theta_{13}=(\widehat{\pp_1,\pp_3})$ and $\phi_{13}=(\widehat{\pp_1+\pp_3,\pp_2+\pp_4})$
[while $\theta=(\widehat{\pp_1,\pp_2})$ and $\phi=(\widehat{\pp_1-\pp_2,\pp_3-\pp_4})$ in \eqqref{OmegaEtheta}].
We switch between the two angular parametrizations using
\begin{eqnarray}
\sin^2\frac{\theta_{13}}{2}&=&\sin^2\frac{\theta}{2} \sin^2\frac{\phi}{2}\\
\cos^2\frac{\theta}{2}&=&\cos^2\frac{\theta_{13}}{2} \cos^2\frac{\phi_{13}}{2}
\end{eqnarray}
and the change of variables
\begin{equation}
\int \frac{\sin\theta_{13}\dd\theta_{13}\dd\phi_{13}}{2\sin\frac{\theta_{13}}{2}}\tilde f(\theta_{13},\phi_{13})=\int  \frac{\sin\theta\dd\theta\dd\phi}{2\cos\frac{\theta}{2}}f(\theta,\phi)
\end{equation}
on an arbitrary function $f(\theta,\phi)=\tilde f(\theta_{13},\phi_{13})$.
Finally the damping rate of the quasiparticles can be expressed as
\begin{equation}
    \Gamma(\pp) = \frac{1}{\tau}\overline{\Gamma}(\varepsilon) \ \text{with} \ \overline{\Gamma}(\varepsilon) = \pi^2+\varepsilon^2
\end{equation}

\textit{Inversion of the collision kernel in an orthogonal basis}---
Following Refs.~\cite{disphydro,Hofmann2023,Hofmann2025} (see also chapter I in \cite{theserepplinger}), we decompose the distributions of total density and polarization $n_\pm=n_\upa\pm n_\dwa$ over an orthogonal basis of polynomials:
\begin{equation}
\begin{pmatrix}
n_+(\pp)\\
n_-(\pp)
\end{pmatrix}=\frac{\partial n_{\rm eq}}{\partial \epsilon}\Bigg\vert_{\epsilon=\epsilon_{\pp\sigma}^0}  \sum_{n,l\in\mathbb{N}}\begin{pmatrix}
U_+ \nu_{n+}^l\\
U_- \nu_{n-}^l
\end{pmatrix} P_l(\cos\theta) Q_n(\epsilon)
\end{equation}
where the orthogonal polynomials $Q_n$ for the energy dependence \cite{disphydro} 
are defined by $Q_0=1$, $Q_1=\epsilon$ and
\begin{eqnarray}
\int^{\infty}_{-\infty}\frac{\textrm{d}\epsilon}{4\cosh^2 \frac{\epsilon}{2}} Q_n(\epsilon)Q_m(\epsilon) = \delta_{n,m}||Q_n||^2 \\
\epsilon Q_n = Q_{n+1}+\xi_n Q_{n-1} \ \text{ with } \ \xi_n = \frac{||Q_n||^2}{||Q_{n-1}||^2}
\end{eqnarray}
The decomposition of $\mathcal{S}$ and $\overline{\Gamma}$ on the $Q_n(\epsilon)$
generates two matrices $\mathcal{S}_{nn'}$ and $\Gamma_{nn'}$:
\bea
\!\Gamma_{nn'}\!\!\!&=&\!\!\!\bb{\pi^2+\xi_{n+1}+\xi_{n}}\delta_{nn'}+\delta_{n-2,n'}+{\delta_{n+2,n'}}\xi_{n+2}\xi_{n+1}\notag\\
\mathcal{S}_{nn'}\!\!\!&=&\!\!\!{2\pi^2\frac{n^2+n-1}{4n^2+4n-3} }\delta_{nn'}\!+\!\frac{\delta_{n-2,n'}}{n(n-1)}\notag
\!+\!\frac{\delta_{n+2,n'}\xi_{n+2}\xi_{n+1}}{(n+2)(n+1)}\\
\eea

The transport equation decomposed over the orthogonal basis then reads
\begin{multline}
\bb{\frac{\omega}{\omega_0}+\frac{\ii}{\omega_0\tau}\mathcal{M}_\pm^l}\vec{\nu}_\pm^l-\frac{l}{2l-1}\bb{1+ \frac{F_{l-1}^\pm}{2l-1}\mathcal{U}_0}\vec{\nu}_\pm^{l-1}\\-\frac{l+1}{2l+3}\bb{1+\frac{ F_{l+1}^\pm}{2l+3}\mathcal{U}_0}\vec{\nu}_\pm^{l+1}=-\delta_{l1}\vec{u}_0
\label{nunl}
\end{multline}
where $\vec{\nu}_\pm^l=(\nu_{n\pm}^l)_{n\in\mathbb{N}}$, $\vec{u}_0=(\delta_{n0})_{n\in\mathbb{N}}$ and $\mathcal{U}_0\vec\nu=(\vec{u}_0\cdot\vec\nu)\vec{u}_0$.
Note the decoupling between total density $n_+$ and polarisation fluctuations $n_-$  which occurs when the equilibrium state is unpolarized.

The dimensionless collision tensor in \eqqref{nunl} is given by
\begin{equation}
\mathcal{M}_{nn'\pm}^l\!=\!
\begin{cases}
\mathcal{M}_{nn'}(\alpha_{n\pm}^l)\equiv \Gamma_{nn'}\!-\!\alpha_{n\pm}^l \mathcal{S}_{nn'} \text{ if } n+n' \text{ is even}\\
0 \text{ else}
\end{cases}\label{Mnl}
\end{equation}
It differs from its value in the case of an isotropic collision probability only via the angular parameters:
\begin{equation}
\alpha_{n\pm}^l=\pm\frac{ 2}{2l+1} \frac{2\Omega_{S\pm}^l-(-1)^n\Omega_{E}^l}{\Omega_{E}^0}
\end{equation}
We have decomposed here $\Omega_E$ and $\Omega_{S\pm}$ from \eqqrefs{OmegaEtheta}{OmegaStheta}
onto Legendre polynomials:
\begin{eqnarray}
\frac{\Omega_{E}^l}{2l+1}&=&\meanvlr{\frac{w_+(\theta,\phi)P_l(\cos\theta)}{2\cos(\theta/2)}}_{\theta,\phi} \label{OmegaE}\\
\frac{\Omega_{S\pm}^l}{2l+1}&=&\meanvlr{\frac{w_\pm(\theta,\phi)P_l(1-2\sin^2\frac{\theta}{2}\sin^2\frac{\phi}{2})}{2\cos(\theta/2)}}_{\theta,\phi}\label{OmegaS} 
\end{eqnarray}
with the angular average $\meanv{f}_{\theta,\phi}=\int_0^\pi \int_0^{2\pi} f(\theta,\phi)\sin\theta\dd\theta\dd\phi/4\pi $. 
In the isotropic case ($W_{\upa\dwa}=\text{cte}$) one has $\Omega_{S+}^l=\Omega_{E}^0$, $\Omega_{S-}^l=0$ and $\Omega_{E}^l=(-1)^l \Omega_{E}^0$.
Note that \eqqref{tau} reads alternatively ${\tau}^{-1}={(m^*)^3T^2}\Omega_E^0/{(2\pi)^3}$.

We close the equations of motion of the 4 conserved
quantities ($U_\pm \nu_{0\pm}^0=-{2\pi^2}\rho_\pm/m^*\pF$, $U_+\nu_{0+}^1=-2\pF m v_\parallel/m^\ast$, $U_+\nu_{1+}^0=-6e/m^\ast T \pF$)
by inverting the transport equation for the corresponding non-conserved 
components (see the second column of Tab.~\ref{tabannexe}).
\setlength\extrarowheight{5pt}
\onecolumngrid
\vspace{\columnsep}
\begin{center}
\begin{table}
\vspace{0.5cm}
\begin{tabular}{c| c |@{\hspace{8pt}}c@{\hspace{8pt}}}
\hline
\hline
{Conserved  quantity} & {Dissipative  quantity} & Collision time to $O(\bar a^2)$ ($\bar a=k_{\rm F}a$)\\
\hline
\makecell{$\nu_{0+}^0, \nu_{0+}^1$\\ $\propto \rho_+,v_\parallel$} & $\nu_{0+}^2$ & \makecell{$\frac{\tau_\eta}{\tau} \!=\!\vec{u}_0 \!\cdot\! \frac{1}{\mathcal{M}^2_+}\!\cdot\!\vec{u}_0 $$\simeq 0.10252+0.08585 \bar a $} \\
\hline
\makecell{$\nu_{1+}^0\propto e$}&$\nu_{1+}^1$&$\frac{\tau_\kappa}{\tau}\!=\!\vec{u}_1 \!\cdot\! \frac{1}{\mathcal{M}^1_+}\!\cdot\!\vec{u}_1 \simeq0.05951+0.02897 \bar a $\\
\hline
$\nu_{0-}^0 \propto \rho_-$&$\nu_{0-}^1$&$\!\!\frac{\tau_D}{\tau}=\vec{u}_0 \!\cdot\! \frac{1}{\mathcal{M}^1_-}\!\cdot\!\vec{u}_0\simeq0.06404+0.02727 \bar a $\\
\hline
\hline
\end{tabular}
\caption{Exact and perturbative expressions of the collision times.\label{tabannexe}}
\end{table}
\end{center}
\twocolumngrid
\vspace{\columnsep}
This requires the inversion of the infinite matrix $\mathcal{M}(\alpha)$ with $\alpha=\alpha_{0+}^2$, $\alpha_{1+}^1$ and 
$\alpha_{0-}^1$ for the viscosity, thermal conductivity and spin diffusivity respectively.
Since the $\alpha'$s are close to their value for
an isotropic collision probability (respectively $2/5$, $2/3$ and $-2/3$),
we evaluate this inverse using the derivative\footnote{We provide via \cite{devviscoMathematica}
the computer algebra program we used to obtain the successive perturbative expansions
of the $\Omega_E^l$, $\Omega_{S\pm}^l$, and $\alpha_{n\pm}^l$. We 
also compute in this way the collision times $\tau_\eta$, $\tau_\kappa$ $\tau_D$ and their derivative
with respect to $\alpha$ from which we finally deduce the transport coefficients.}: $\dd\mathcal{M}^{-1}/\dd\alpha =\mathcal{M}^{-1}\mathcal{S}\mathcal{M}^{-1}$.

\section{Landau-Salpeter equation on the forward scattering amplitudes}
\label{LandauSalpeter}

Using $\ket{i}=\ket{f}=\ket{\{n_{\pp\sigma}\}}$ in \eqqref{Aif},
we obtain the energy shift of $\ket{i}$ to second order in $\hat V$ \cite{LandauLipschitzVol9}:
\begin{multline}
\bra{\{n_{\pp\sigma}\}}\hat H\ket{\{n_{\pp\sigma}\}}=E_i^0+\frac{g_0}{L^3}\sum_{\pp\pp'\in\mathcal{D}} n_{\pp\upa} n_{\pp'\dwa}\\+\frac{g_0^2}{L^6}\sum_{\pp_1\pp_2\pp_3\pp_4\in\mathcal{D}}\frac{(1-n_{\pp_1\upa})(1-n_{\pp_2\dwa})n_{\pp_3\dwa} n_{\pp_4\upa}}{(\pp_3^2+\pp_4^2-\pp_1^2-\pp_2^2)/2m}\delta_{\pp_1+\pp_2}^{\pp_3+\pp_4}
\end{multline}
Applying the discreet derivative rule \eqqref{fpp} to this, we express the interaction functions
as
\begin{multline}
{f_{\upa\dwa}(\pp,\pp')}={g_0}+\\\frac{g_0^2}{L^3}\sum_{\pp_a\pp_b\neq\pp,\pp'}\bbcrol{\frac{(1-n_{\pp_a}^0)(1-n_{\pp_b}^0)-n_{\pp_a}^0n_{\pp_b}^0}{(p^2+p'^2-p_a^2-p_b^2)/2m}\delta_{\pp+\pp'}^{\pp_a+\pp_b}}\\\bbcror{+\frac{n_{\pp_a}^0-n_{\pp_b}^0}{(p^2+p_b^2-p'^2-p_a^2)/2m}\delta_{\pp+\pp_b}^{\pp_a+\pp'}}\label{fupdwcachee}
\end{multline}
\begin{multline}
\!\!\! {f_{\sigma\sigma}(\pp,\pp')}=\frac{g_0^2}{L^3}\!\!\!\!\!\sum_{\substack{\pp_a,\pp_b\neq\pp,\pp'}}\frac{(n_{\pp_a}^0-n_{\pp_b}^0)\delta_{\pp+\pp_b}^{\pp'+\pp_a} }{(p^2+p_b^2-p'^2-p_a^2)/2m}\label{fupupcache}
\end{multline}

Expression \eqref{fupdwcachee} of $f_{\upa\dwa}$ shows a UV divergence when the lattice spacing $l$ tends to 0. This
is regularized by eliminating $g_0$ in favor of $a$. Expanding \eqref{gg0} for $a\to0$ at fixed $l$, we obtain
\begin{equation}
g_0=g+g^2 \mathcal{P}\!\!\!\!\!\!\!\!\underset{[-\pi/l,\pi/l)^3}\int\!\!\!\!\!\!\!\!\frac{\dd^3 p_a \dd^3 p_b}{(2\pi)^3}\frac{\delta(\pp_1+\pp_2-\pp_a-\pp_b)}{(p_a^2+p_b^2-p_1^2-p_2^2)/2m}
+O(g^3)
\end{equation}
This relation is valid in the continuous space limit $l\to0$ (taken after $a\to0$) where the integral on the right-hand side
is independent of $\pp_1$, $\pp_2$ and can be view as a rewriting of the one appearing in \eqqref{gg0}.
In \eqqref{Aupdw}, the contribution of diagram I is regularized in the same way as $f_{\upa\dwa}$.
To obtain  \eqqrefs{fupdw}{fupup} in the main text, there remains to compute the regularized integrations over the intermediate momenta $\pp_a$ and $\pp_b$.

As explained in the main text, the forward scattering amplitudes $\mathcal{A}_{\sigma\sigma'}(\pp,\pp')\equiv\mathcal{A}_{\sigma\sigma'}(\pp,\pp'|\pp',\pp)$
can be deduced from $f_{\sigma\sigma'}$ through 
the Landau-Salpeter equation. In our notations, Eq.~(33) from Ref.~\cite{Landau1959} becomes:
\begin{multline}
\mathcal{A}_{\sigma\sigma'}(\pp,\pp')=f_{\sigma\sigma'}(\pp,\pp')\\-\frac{\pF^2}{(2\pi)^3\vF}\sum_{\sigma''=\uparrow,\downarrow}\int\dd\Omega'' f_{\sigma,\sigma''}(\pp,\pp'') \mathcal{A}_{\sigma''\sigma'}(\pp'',\pp')
\end{multline}
where $\dd\Omega''=\sin\theta''\dd\theta''\dd\phi''$ is the solid angle locating $\pp''$. To second order in $g$, this is no longer an integral equation and this evaluates in
\bea
\mathcal{A}_{\uparrow\downarrow}(\pp,\pp')&=&f_{\uparrow\downarrow}(\pp,\pp')+O(g^3)\\
\mathcal{A}_{\sigma\sigma}(\pp,\pp')&=&f_{\sigma\sigma}(\pp,\pp')-\frac{2\kF a}{\pi}g+O(g^3)
\eea
which recovers \eqqrefs{Aupdwforward}{Aupupforward}. The Landau-Salpeter equation is thus an elegant way of deriving a limiting value of the off-diagonal elements
$\bra{f}\hat H\ket{i}$ from the knowledge of the diagonal elements $\bra{i}\hat H\ket{i}$ only.

\sloppy
\bibliography{../HKLatex/biblio}

\end{document}